\documentclass[letterpaper]{emulateapj}

\slugcomment{ApJ accepted}

\shorttitle{Contamination by Type~I\lowercase{bc} SNe}
\shortauthors{Homeier}

\begin{document}


\title{The Effect of Type I\lowercase{bc} Contamination in Cosmological Supernova Samples}


\author{N. L. Homeier}
\affil{Department of Physics and Astronomy, Johns Hopkins University, 3400 North Charles Street, Baltimore, Maryland, 21218-2686}



\begin{abstract}
We explore the effect of contamination of intermediate redshift
Type~Ia supernova samples by Type~Ibc supernovae. 
Simulating observed samples of Ia and mixed 
Ibc/Ia populations at a range of redshifts for an underlying
cosmological concordance model ($\Omega_{m}=0.27$, $\Omega_{\Lambda}=0.73$),
we find that even small contamination levels, $2-5$\%
may bias the derived $\Omega_{\Lambda}$ and $\Omega_{m}$ towards
larger values. We thus
emphasize the need for clean samples of Type~Ia SNe for accurate 
measurements of the cosmological parameters.
We also simulate a SN sample similar to the fiducial SNAP detected 
distribution
(Kim et~al. 2004), but include Ibc contamination. For this distribution 
we illustrate the effect of Ibc 
contamination on the distance modulus vs. 
redshift diagram for low and high precision measurements.

%As Type~Ibc SN come from the most massive stars, their rate of 
%occurence depends on the star formation rate in the previous 10~Myr. 
%Thus, we should expect that the Ibc/Ia SN rate approaches or exceeds
%1.0 at $z=1$. This motivated us to investigate what effect Type~Ibc SN
%would have on assumed Ia samples, and thus on the cosmological parameters
%probed.We assume an
%apparent magnitude limit of $m=25$ to derive the change in the measured
%distance modulus as a function of redshift for Ibc/Ia ratios of 0.1, 0.5,
%and 1.0. We find that contamination
%by Type~Ibc SN in an assumed Ia sample has no significant effect on the
%{\it median} measured distance modulus, even for equal numbers of Type Ibc
%and Ia SN. However, the {\it mean} measured distance modulus for a mixed 
%sample changes significantly; the sample appears fainter at
%at $z=0.5-0.7$ (depending on the cosmological model and 
%the Ibc magnitude distribution), and a brighter at higher redshifts. 
%The inflection point is set by the apparent 
%magnitude limit, the cosmological model, and the magnitude distribution
%of Type~Ibc and Ia SN.

\end{abstract}




\section{Introduction}

Type~Ia supernovae (SNe) are considered astronomical standard candles and 
have been used to measure the geometry and dynamics of the Universe 
\citep{Riessetal98,Perletal99}. 
This exceedingly difficult measurement relies on observations of local 
supernovae samples to calibrate relationships between absolute peak magnitude
and more easily obtained observables such as light-curve width, color, etc. 
(see \citet{Leib01} for a review). These calibrations
are then applied to the observables in higher redshift samples to derive 
distance moduli.
Due to the extreme faintness of intermediate to high redshift SN and the
shifting of identifying spectral features outside the optical window,
spectroscopic classification becomes difficult,
and photometry must often suffice. Fortunately, there is a 
clear observational 
difference between hydrogen-rich (Type~II) and hydrogen-poor (Ia, Ib, and Ic) 
SNe, namely, the UV deficit (UV radiation is blocked in hydrogen-poor
SN explosions by metal lines). This property can be used to distinguish 
between Type~I and II SNe explosions. To date there is no  
photometric technique to distinguish between Type~Ia and Ibc SN (although see
Gal-Yam et~al., 2004), and one
must perform a spectroscopic classification. The characteristic spectroscopic 
difference between Type~Ia and Ibc SNe is the
deep absorption trough at 6140\AA~ in Ia spectra, which is due to the 
blueshifted \ion{Si}{2} $\lambda\lambda 6347, 6371$\AA~ feature. Unfortunately
this is redshifted out of the optical passband for $z>0.5$, making 
classification of Type~I SNe difficult. Alternative spectral
discriminants between Ia and Ibc SNe are \ion{Si}{2} $\lambda 4130$ 
(e.g. Coil et~al. 2000), \ion{Fe}{2} $\lambda 4555$ and \ion{Mg}{2} 
$\lambda 4481$ \citep{Barrisetal04}. Spectroscopic identifications 
are made using template-matching, but no statistical information is
given about misidentifications with this method \citep{Tonetal03,Retal04b,Barrisetal04}.

%Type~Ib and Ic supernovae come from massive stars (M$> 20-30$~M$_{\odot}$)
%after they have lost their hydrogen envelopes through
%stellar winds and mass ejections (likely a combination of the two)
%throughout their lifetime. Type~II and Ibc are similar in that they 
%involve core-collapse of a massive star, whereas the progenitors of 
%Type~Ia SN are low mass stars. The two leading theoretical progenitors for 
%Type~Ia SN are singly-degenerate (WD+RG)and doubly-degenerate (WD+WD) models.
%The ability to photometrically differentiate between Type~Ia and
%Type~II SN appears secure, however, it is extremely difficult to 
%differentiate between Type~Ia and Ibc SN without moderate S/N spectra.
%The largest spectroscopic difference between Type~Ia and Ibc SN is the
%deep absorption trough at $6140\AA$ in Ia spectra, which is due to the 
%blueshifted \ion{Si}{2} $\lambda\lambda 6347, 6371$ feature. Unfortunately
%this is is redshifted out of the optical passband for $z>0.5$, making 
%%classification of Type I SN extremely difficult. 

Assuming the same relative Ia and Ibc rates in nearby SNe samples,
 Riess~et.~al~(2004) and Barris~et~al.~(2004)
reasoned that the rate of contamination from Type~Ibc SN in 
high redshift samples should be less than $\sim10$\%. However,  
the star formation rate increases rapidly 
at up to $z\sim 1-2$, and this is likely to have a significant 
effect on the relative rates of SN types. As Type~Ibc SNe come from
massive stars (M~$> 20-30$~M$_{\odot}$), their rate depends on the star 
formation rate in the previous 10~Myr. It is well established that the 
global star formation rate increases at 
higher redshifts. Hence, assuming a delay for Ia explosions,
we expect that the rate of Type~Ibc SNe increases 
over the short term, and that the contribution from 
Ibc SNe becomes more important at higher redshift. 
Motivated by this, we investigate here what effect contamination by Type~Ibc
SNe has on cosmological parameters derived from Ia SNe samples. 
The optimal way to determine the systematic effect in a 
particular survey is through simulations using the
diagnostic machinery associated with said survey. However,
the general nature of the effect can be investigated
robustly with simulations such as the one we present here.

\section{The global star formation rate}

The global star formation rate peaked somewhere between $z\sim1-2$ and has 
been declining to the present day \citep{Letal96, Madetal98, Steidetal99, 
Thometal01, Bouwetal03}. Between the local Universe and z$\approx 1.5$, the 
star formation rate increased to $10-20$ times the rate in the current
epoch. This significant increase in star formation rate should
have a substantial effect on the relative rates of SN types. If the rates are
about 1 Type~Ibc SN for every 9 of Type~Ia, when
the star formation rate was 10 times its present value, the rates of
Type~Ibc and Ia SN should be about equal; for higher star formation rates,
Type~Ibc SN will outnumber those of Type~Ia. Here we assume that the rate 
of Type~Ibc SN explosions does not evolve with metallicity, therefore it 
depends only on the number of massive stars formed in the last $1-10$~Myr, 
and thus scales linearly with the recent star formation rate. We also
assume that the rate of Type~Ia SN should be relatively constant after 
the ``time delay'' until the first explosion, which is somewhat speculative. 
It is possible that the rate of Type~Ia SNe correlates with 
the star formation rate to some degree, but in any case to a lesser degree 
than the core-collapse Type~Ibc SNe. Whatever the actual degree,
an increase in the global star formation rate will affect the relative
ratio of Type Ia to Ibc SNe observed at that epoch, which leads us to the 
present investigation.
Important constraints on the
time delay for SNe~Ia explosion are now being extracted from
intermediate redshift SNe samples \citep{Dahlenetal04,Strolgeretal04}.
However, it is important to note 
that the general conclusions presented here
{\it do not depend} on the Type~Ia vs. Ibc rate with star formation rate,
only that an increasing Ibc/Ia rate with redshift provided the motivation
for this study.

\section{Estimating the Effect of I\lowercase{bc} contamination}

\subsection{Absolute Magnitude Distributions}
\label{abmag}

%%\clearpage
\begin{figure}
  \includegraphics[scale=0.5]{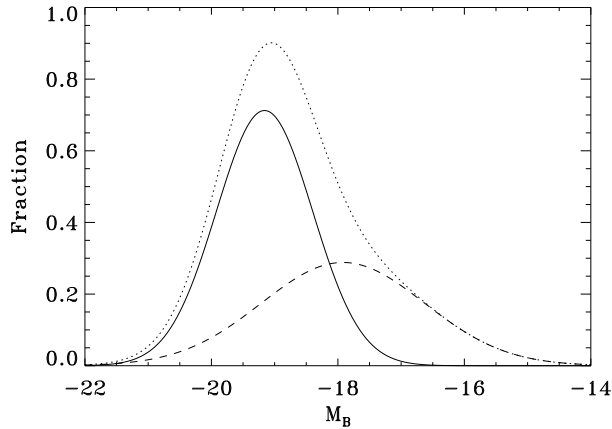}
  \caption{Magnitude distribution function for Type~Ia and Ibc SNe assuming
equal numbers of each and a single gaussian distribution for Type~Ibc SNe. 
\label{mag1}}
\end{figure}

%\clearpage
\begin{figure}
  \includegraphics[scale=0.5]{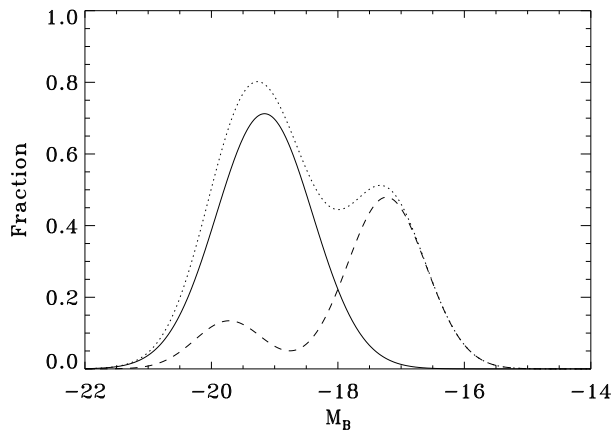}
  \caption{Magnitude distribution function for Type~Ia and Ibc SN assuming
equal numbers of each and a double gaussian distribution with ``bright'' 
and ``normal'' Type~Ibc SN. 
\label{mag2}}
\end{figure}
%\clearpage

We use the observed distribution of uncorrected absolute magnitudes,
from the \citet{Richetal02} study of the Asiago Supernova Catalog.
Pertinent here
are the 111 Type~Ia SN, 5 luminous Type~Ibc SN, and 13 ``normal'' Type~Ibc SN
in the catalog. We will use their single gaussian fit to the {\it uncorrected}
absolute magnitudes of the 111 Type~Ia SN ($M_{B}=-19.16, \sigma = 0.76$) and 
their single gaussian fit to the Type~Ibc SNe 
($M_{B}=-18.92, \sigma=1.29$); these are shown in Figure~\ref{mag1}. 
The number of Type~Ibc SNe in the sample
is small enough that a two-component model for the distribution of 
magnitudes is equally valid, with 
$M_{B1}=-17.23, \sigma_{1}=0.62$; $M_{B2}=-19.72,\sigma_{2}=0.54$, 
and a weight$=0.28$. We run models with both the one and two gaussian
distributions for Ibc magnitudes.

\subsection{Method}

%\clearpage
\begin{figure*}
\begin{center}
  \includegraphics[width=14cm]{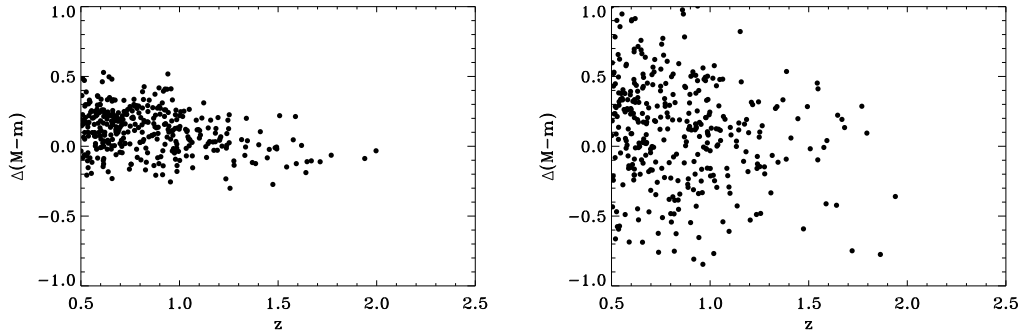}
  \caption{An example of detected SN redshift distributions 
(1000 input, $320-360$ detected) SNe and varying the ''observational scatter''
from (a) 0.05, where one is dominated by Ia magnitude corrections, $\sigma = 0.14$ and 
(b) 0.35, where the observational scatter dominates. Input $(\Omega_{M},\Omega_{\Lambda})=(0.27,0.73)$.   
\label{}}
\end{center}
\end{figure*}
%\clearpage

We use a Monte Carlo simulation that creates a random redshift array
from $z=0.5-3$. Absolute magnitudes are assigned 
from the Type Ia and Ibc absolute magnitude
distributions (described in the preceding section). We then 
include ``observational scatter'' by
adding or subtracting flux from each object according
to a gaussian distribution with $\sigma=0.35$ magnitudes. This is 
slightly larger than the mean error of SNe between $z=0.5-2$ in
the \citet{Tonetal03} sample (32 objects, mean error$=0.30$).

We ``observe'' objects 
by calculating the luminosity distance for each object at its 
individual redshift, and ``detecting'' those with absolute magnitudes
brighter than the detection limit, which we take as $m_{B}=25$. 
This should
be roughly correct for redshifts up to $z=1$. Beyond this, rest-frame
$B$ corresponds most closely to $J$, and $M_{J}=24$ is a more
realistic limit. However, including this effect does not significantly
affect the results in what follows.

The Type Ia SNe that are 
``detected'' are re-sampled to a Gaussian distribution with a center
at what we take as the 
corrected Type Ia SNe absolute magnitude (-19.41; Leibundgut 2001), 
but with the 
dispersion equal to quadrature addition of the corrected dispersion 
of a typical low-redshift SNe sample
(0.14; Barris et~al. 2004) and an additional
observational scatter. This simulates the
corrections made to the Type Ia absolute magnitudes based on
light curve width and color, but with a larger scatter to 
simulate observational error. 
Distance moduli are calculated for each object by assuming each
has the absolute magnitude of a corrected Type~Ia SN. Then the difference
between these distance moduli and the distance moduli for an 
empty universe is calculated.

High redshift supernovae surveys utilize an
analysis method based on the 'Multiwavelength' or 'Multicolor 
Light-curve Shape Method' \citep{Retal04b,Barrisetal04}.
This method uses a set of Ia lightcurves as templates for
training, then concurrently fits the distance modulus, 
extinction, typically expressed in units of $A_{V}$, and $\Delta$,
a parameter which describes the lightcurve shape and
corresponds to the difference in absolute magnitude of 
the SN and a fiducial SN Ia. 

The dominant correction comes from the correlation
between light curve shape and peak magnitude,
in the sense that fainter Ia SNe decline in 
luminosity faster than brighter ones. In order to
discuss this quantitatively, we consider here the
decline parameter, $\Delta m_{15}$, which is
the difference between the peak
magnitude and the magnitude at 15 days after maximum (e.g. 
Phillips 1993). The decline parameter varies with the observed restframe
wavelength, and depends on K-corrections, 
extinction, and redshift (time dilation) \citep{NKP02}. The
goal here is not to reproduce a particular survey
and its associated completeness and bias, but to investigate
any possible systematic effect a contaminated SNe sample 
has on the derived cosmological parameters. Therefore, in this study 
we assume that the corrections (K,
extinction, and time dilation) are accurately performed
and the errors associated with these corrections are
contained in the observational scatter parameter.

To estimate typical modifications to Ibc magnitudes,
we use the relation from \citet{Hetal96}
for the B-band, $M_{MAX}=a + b[\Delta m_{15}(B)-1.1]$,
where $b=0.784$, and we set $a$ to the value we use for
corrected peak Ia magnitudes, -19.41.
We stress that the particular values for
$M_{MAX}$, dispersions, etc. that we use are not significant, only
that we consistently use the same values throughout 
the simulation.
Then we need typical decline parameters
for Ibc SNe.
Most published lightcurves for Ib and Ic SNe are not 
well-sampled in the period $\pm 30$ days from maximum,
but focus on the long-term behavior of the lightcurve. 
We have collected a sample of 5 Ibc lightcurves with
measurable
$\Delta m_{15}$;
these are summarized in Table~\ref{tab2}.
We also note that we have not listed the Ib SN~1983N 
\citep{Cloetal96} and Ic SN~1983V 
\citep{Cloetal97} that have been shown to have similar
lightcurves to SN~1993J, although the number of
photometric points in the first $\pm 30$ days is
small.

%\subsection{SN1990B}

%With the B and R photometry published in \citet{Cloetal01},
%we use a polynomial fit to measure $\Delta m_{15}(B) = 0.96$ 
%and $\Delta m_{15}(R) = 1.14$. These values are
%uncertain because SN1990B was not observed
%before maximum, i.e. only a decline was observed.

%\subsection{SN1994I}

%For the Ic SN 1994I in M51, \citet{Retal96} measured decline
%parameters $\Delta m_{15}(B) = 2.07\pm0.03$
%and $\Delta m_{15}(I) = 1.08\pm0.02$.

%\subsection{SNe 1998dt, 1999di, 1999dn}

%Using photometry for SNe published in \citet{Metal01}, 
%we use simple third to fourth order polynomial fits
%to estimate decline parameters. We find
%$\Delta m_{15}(R) = 0.2-0.3$ for 1998dt, 
%$\Delta m_{15}(R) = 0.40$ for 1999di, and $\Delta m_{15}(R) = 0.46$
%for 1999dn. It should be noted that the magnitudes for 1998dt and 1999di
%are unfiltered from the KAIT CCD system, and approximate $R$.

\subsection{Decline Parameters}

Decline parameters for these 5 Ibc SNe span a wide range of values.
In light of this, we select reasonable 
ranges for the decline parameter considering the sample
of Ia SNe presented in \citet{Hetal96}. We emphasize that
this is valid even if decline parameters for Ibc
SNe are distributed differently than Ia decline 
parameters, as Ibc SNe that are not weeded out will be
fit to Ia templates. We explore four
treatments assuming: (1) no modifications, (2) all Ibc SNe are matched to
templates with small decline parameters (mean=0.9, dispersion=0.1), 
(3) all Ibc SNe
are matched to templates with large decline parameters
(mean=1.5, dispersion=0.3),
and (4) Ibc SNe are matched with decline parameters
with a mean of 1.1 and a gaussian dispersion of 0.3.
This leads to mean assumed absolute magnitudes of (1) $-19.41$
(2) $-19.57$, (3) $-19.10$, and (4) $-19.41$ (same as (1)).

%\clearpage
\begin{table}
\begin{tabular}{lcllcc}\hline
SN & Type & $\Delta M_{15}$ & M$_{B,max}$ & $B_{max}-V_{max}$ & Ref \\\hline
1990B     & Ic     & 0.964 (B)  & $-$ & $-$ & (1), (2)\\
          &        & 1.14  (R)  &  &  &        \\
1992ar    & Ic     &                     &  $-19.7\pm 0.5$  & $0.4\pm 0.2$ & (3)\\ 
1994I     & Ic     & $2.07\pm 0.03$ (B)   & $-17.68\pm 0.73$ & 0.15$^{*}$& (4) \\
1993J     & IIb    & 2.5 (B)                    &  $-17.3\pm 0.6$ & 0.18$^{*}$ & (6)\\
          &        & 0.95 (V)                   &                 &            & (6)\\
          &        & 0.8 (R)                    &                 &            & (6)\\
1998dt    & Ib     & 0.2-0.3 (R)         & $-$ & $-$ & (2) \\
1999di    & Ib     & 0.4 (R)             & $-$ & $-$ & (2) \\
1999dn    & Ib     & 0.5 (R)             & $-$ & $-$ & (2) \\
1999ex    & Ic     &                     & $-17.4$  & 0.42 & (7) \\
%1983N     & Ib     & 1.4 (B)             &  $-17.4\pm 0.3$ & ---- &   \\
%1983V     & Ic     & 0.6 (B)             &  $-18.27\pm 0.44$ & 0.34 &  \\
%1984L     & Ib     & 0.7 (B)             &  $-17.4$
%1985f     &        &                     &  $-18.4\pm 0.7$  &     & \\ 
1999bw    & Ic     & 1.0 (B)             &  $-18.88$        & 0.4 &  (8)\\\hline
%2002ap    & Ic     & 1.3 (B)             &  $-16.3$         & 0.6 &  \\
\footnotetext{*Corrected for extinction, 
(1) Clocchiatti et al. (2001), 
(2) Matheson et al. (2001), 
(3) Clocchiatti et al. (2000), 
(4) Clocchiatti et al. (1996)
(5) Richmond et al. (1996), 
(6) Richmond et al. (1994), 
(7) Stritzinger et al. (2002)
(8) Galama et~al. (1998)}
\end{tabular}
\label{tab2}
\end{table}
%\clearpage

We compare
these simulated data to a grid of
models with a range of $\Omega_{m}$ and $\Omega_{\lambda}$ values, 
and calculate the corresponding grid of $\chi^{2}$ values.
Probability curves are calculated from this $\chi^{2}$ grid 
following the method of \citet{Tonetal03}. That is, the minimum
in the $\chi^{2}$ array is found, and a grid of probabilities
is calculated where prob=exp($-0.5\times(\chi^{2}-min(\chi^{2}))$.
The levels of 68\%, 95\%, and 99.5\% are found by integrating (summing) the 
probability array and identifying where 68\%, 95\%, and 99.5\%
of the total probability are located. This is only valid when the 
minimum reduced $\chi^{2}$ is close to 1 (as it is in the SNe sample
analyzed in \citet{Tonetal03}).

\section{Results}

\subsection{Malmquist Bias with Ibc SNe}

\begin{figure}
  \includegraphics[scale=0.45]{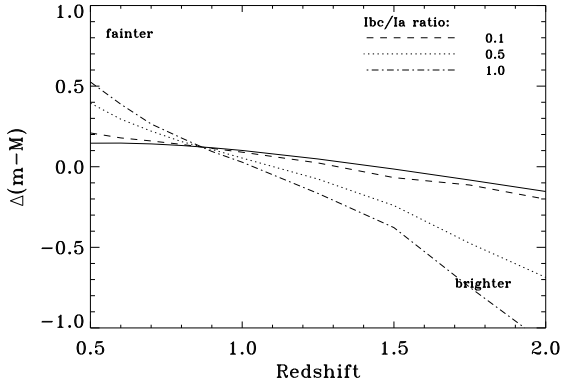}
  \caption{Change in the mean measured distance modulus for Ibc/Ia ratios
of 0.1, 0.5, and 1.0. This is shown only for purposes of illustration.
\label{delta_mean}}
\end{figure}

In Figure~\ref{delta_mean} we show the change in the measured mean
distance modulus (compared to an empty universe or eternally coasting
model) from our simulation for {\it input} Ibc/Ia ratios of 0.1, 0.5, 
and 1.0 and an underlying concordance model with 
$(\Omega_{m},\Omega_{\Lambda})=(0.27,0.73)$. An interesting effect
can be seen, which is solely due to the observational apparent
magnitude limit. At redshifts less than $\sim0.8$, the mean distance
modulus is fainter than the pure Ia case, and at redshifts greater than 
this it is brighter. This is because of the luminosity distribution 
of Type~Ibc SNe, with the majority being fainter than Type~Ia SNe.
At lower redshifts, Ibc contamination pushes the mean fainter, while
at higher redshifts, these fainter SNe are below the detection limit, 
and only the brightest SNe are detected, pushing the
mean to apparently brighter magnitudes. The transition point is set by the 
apparent magnitude limit, here taken as $m_{B}=25.$, and moves to
higher redshift at fainter apparent magnitude limits, and vice versa.
This is shown only for illustrative purposes, and no conclusions are
derived from this diagram. 

%\clearpage
%\clearpage

\begin{figure*}
\begin{center}
\includegraphics[scale=0.68]{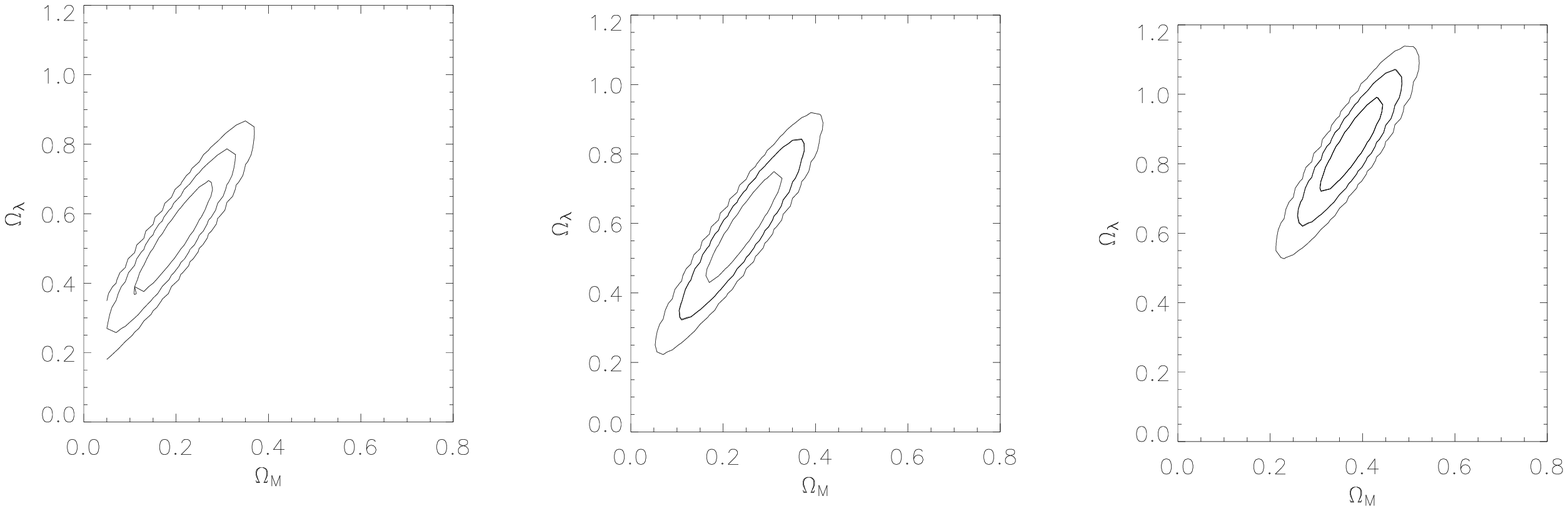}
\caption{Input of 2000 Ia uniformly and randomly distributed from z$=0.5-2$ and a single gaussian distribution of
(b) 100 Ibc and (c) 200 Ibc. 
(a) 724 detected Type Ia SNe. The minimum reduced $\chi^{2}$ is 1.03.
(b) 724 detected Type Ia SNe, 11 Ibc SNe. The minimum
reduced $\chi^{2}$ is 1.13. (c) 724 detected Type Ia SNe, 26 detected Ibc SNe.
The minimum reduced $\chi^{2}$ is 1.36.
\label{omol}}
\end{center}
\end{figure*}
%\clearpage

\subsection{Stochasticity}

The number of detected SNe affects the spacing of the
probability curves, but more importantly, it affects the 
variation in the location of the $\chi^{2}$ minimum from run to run.
As expected, with smaller numbers of detected SNe, the curves 
are unstable, and ''jump around'' from run to run. As the number 
of input SNe is increased,
and thus the number of detected SNe also increases, the position of
the curves is more stable. 

As an example of this, for 100 trials using 500 Type~Ia SNe
uniformly and randomly distributed between z$=0.5-2.0$, with a mean of 
190.04 detected 
SNe, the mean $\Omega_{m}, \Omega_{\Lambda}$ values at the location of the
$\chi^{2}$ minima were 0.269 and 0.733, with standard deviations of 0.103 
and 0.181. The mean minimum reduced $\chi^{2}$ was 1.01, with a standard deviation 
of 0.11.
Inputing 5000 Type~Ia SNe in 100 trials with a mean of 1884.07 detected SNe
decreased the standard deviations to 0.034 and 0.064, with mean $\Omega_{m}, 
\Omega_{\Lambda}$ values of 0.27 and 0.73. The mean minimum $\chi^{2}$/N 
was 1.007, with a standard deviation of 0.031. 

%If we change the redshift range to z$=0.5-3.0$ and
%input 800 Ia SNe, with a mean of 180.03 of detected SNe
%the mean $\Omega_{m}, \Omega_{\Lambda}$ values at the location of the
%$\chi^{2}$ minima were 0.245 and 0.693, with standard deviations of 0.092 and 0.165. 
%The mean minimum $\chi^{2}$/N was 1.009, with a standard deviation 
%of 0.089.

We investigated what effect Type~Ibc SNe would have on the derived 
probability curves by assuming values of 0.27 and 0.73 for $\Omega_{m}$
and $\Omega_{\Lambda}$ (the 'cosmological concordance' model)
and distributing 2000 Type~Ia SNe uniformly and randomly between $z=0.5-2$.
In one run of the simulation, 724 of the 2000 were ''detected'', 
and the probability curves corresponding to the $\chi^{2}$ analysis
are shown in Figure~\ref{omol}(a). Using these same 2000/724 Type~Ia 
SNe, we first 
added 100 Type~Ibc SNe with magnitudes distributed randomly according
to the unimodal distribution presented in \S~\ref{abmag}. 
11 of these were ``detected'', and the result
of this simulation is shown in Figure~\ref{omol}(b). In 
Figure~\ref{omol}(c) we show the result of using 200
Type~Ibc SNe (input ratio of 10\%), of which 26 were detected.
The reduced $\chi^{2}$ increased from 0.97 for the pure Ia sample, 
to 1.13 for the sample used in Figure~\ref{omol}(b), to 1.36
at a contamination level of only 5\%, shown in Figure~\ref{omol}(c).

In these simulations, relatively low contamination rates of 26/742=4\%
significantly affect the location of the probability curves in
the $\Omega_{M}, \Omega_{\Lambda}$ plane, moving both $\Omega_{M}, \Omega_{\Lambda}$
to higher values. However, as the 
fraction of detected Type~Ibc SNe increases, the minimum 
reduced $\chi^{2}$ value also increases. At some point, this
minimum reduced $\chi^{2}$ value should be large enough that
one would correctly conclude that the model is not a good fit,
and the conversion from the $\chi^{2}$ grid to probability curves 
would not be, and should not be, attempted without first renormalizing 
the errors. Then the probability contours become larger, and should
encompass the correct $\Omega_{M}, \Omega_{\Lambda}$ values.
However, in this renormalization it is implicit that the
model is a good fit, and the random errors are underestimated. In this 
example, this is incorrect.

The spacing of the probability curves changes with the value of the
observational scatter and the total number of SNe detected. 
But the stability of the location of the center of the curves depends on the
number of detected SNe, varying stochastically. 
We find that for large samples with 
relatively accurate measurements, even a small 
contamination ($2-3$\%) of Type~Ibc SNe can significantly bias the
result. 
In 100 trials with an input of 5000 SNe uniformly and randomly distributed
between z$=0.5-2.0$,  
an observational scatter of 0.35, an input 
Ibc/Ia ratio of 0.1, and a bimodal magnitude distribution
(resulting in a mean of 1891.05 Type~Ia and 54.61 Type~Ibc
SNe detected), 
the mean $\Omega_{m}, \Omega_{\Lambda}$ 
values for the location of the $\chi^{2}$ minimum were 0.334 and 0.826,
with standard deviations of 0.038 and 0.069. 
The minimum reduced $\chi^{2}$ was 1.156, with a standard
deviation of 0.048. In this case, the underlying values of 
$\Omega_{m}=0.23, \Omega_{\Lambda}=0.73$ are outside the  $1\sigma$ curves,
but inside the $2\sigma$ curves,
68\% of the time.

It is worth noting that the minimum $\chi^{2}$ is at 
$(\Omega_{m},\Omega_{\Lambda})\sim(1.3,0.6)$ for the \citet{Tonetal03}
sample, before applying external constraints. In fact, without
external constraints, the ``concordance
model'' values are outside the 1$\sigma$ contours. 
However, in \citet{Retal04b} and \citet{Barrisetal04}, the
error contours are significantly larger, and encompass the
concordance model values for the density parameters.

\subsection{Unimodality vs. Bimodality}

We also compared the effects of 
single and double gaussian Ibc magnitude distributions. 
These results are summarized in Table~\ref{tab1}. With 
a single gaussian distribution, stochastic effects are
more important, as illustrated by the larger standard 
deviations for the mean recovered $\Omega_{M}$, $\Omega_{\Lambda}$,
and minimum $\chi^{2}/$N. In terms of systematic error in
the recovered cosmological parameters, 
if the double gaussian distribution for Ibc magnitudes is 
more accurate in describing the underlying distribution,
contamination by Ibc SNe will bias the results less severely
than the single gaussian case.

%With 100 trials of 100 Type Ia SN, the mean recovered $\Omega_{m}, 
%\Omega_{\Lambda}$ is 0.27 and 0.75, with standard deviations of 0.16 and
%0.30. This standard deviation is quite high.
%With 100 trials of 90 Type Ia SN and 10 Type Ibc, the mean recovered 
%$\Omega_{m}, 
%\Omega_{\Lambda}$ is 0.35 and 0.87, with standard deviations of 0.22
%and 0.41
%With 100 trials of 180 Type Ia SN and 20 Type Ibc, the mean recovered
%$\Omega_{m}, \Omega_{\Lambda}$ is 0.34 and 0.86, with standard deviations
%of 0.14 and 0.28.
%With 100 trials of 200 Type Ia SN, the mean recovered $\Omega_{m}, 
%\Omega_{\Lambda}$ is 0.27 and 0.72, with standard deviations of 
%0.14 and 0.27

%\begin{figure}
%\includegraphics[scale=0.5]{../sntypes/omol_500ain.eps}
%\caption{An example with a smaller sample of detected SNe.
%(a) 117 detected Type Ia SNe. The minimum
%reduced $\chi^{2}$ is 0.95.  (b) 117 detected Type Ia SNe, 5 Ibc SNe.  
%The minimum reduced $\chi^{2}$ is 1.44
%\label{smallsamp}}
%\end{figure}
%\clearpage
\begin{table}[width=7cm]
\caption{A comparison of the magnitude distributions for Ibc supernovae. Results from 100 trials.}
\begin{tabular}{lcc}\hline
                  & single gaussian & double gaussian \\\hline
mean $\Omega_{M}$       & $0.409\pm 0.126$ & $0.344\pm 0.094$ \\
mean $\Omega_{\Lambda}$ & $0.974\pm 0.243$ & $0.839\pm 0.176$ \\
mean minimum $\chi^{2}/$N & $1.393\pm0.242$ & $1.225\pm0.110$ \\
mean number of SNe Ia detected  & 228.77  & 226.78  \\  
mean number of SNe Ibc detected &  12.46  &  10.26  \\ 
\label{tab1}
\end{tabular}
\end{table}
%\clearpage
\subsection{Effect of Decline Parameters}

%\clearpage
\begin{figure*}
\begin{center}
\includegraphics[width=14cm]{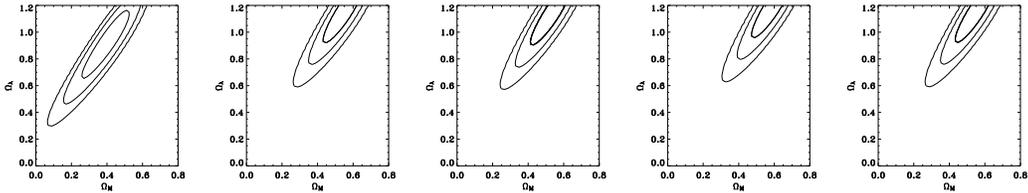}
\caption{See text for details. From left to right: (a) 231 Ia SNe, (b) 231~Ia, 15~Ibc with no modification 
to the Ibc magnitudes, (c) 231~Ia, 15~Ibc with modification with small decline parameters, (d) 231~Ia, 15~Ibc 
with modification with large decline parameters, (e) 231~Ia, 15~Ibc with modification with medium decline 
parameters. There is no significant difference between scenarios with this range of Ibc magnitude modifications. 
  \label{declinecomp}}
\end{center}
\end{figure*}
%\clearpage

We explore four treatments assuming: (1) no modifications to Ibc magnitudes, 
(2) all Ibc SNe are matched to
templates with small decline parameters (mean=0.9, dispersion=0.1), 
(3) all Ibc SNe
are matched to templates with large decline parameters
(mean=1.5, dispersion=0.3),
and (4) Ibc SNe are matched with decline parameters
with a mean of 1.1 and a gaussian dispersion of 0.3.
This leads to mean assumed absolute magnitudes of (1) $-19.41$
(2) $-19.57$, (3) $-19.10$, and (4) $-19.41$ (same as (1)).

In Figure~\ref{declinecomp} we compare the results for
a single case for these four treatments.
The number of detected SNe was 231 Ia and 15 Ibc.
Assuming Ibc SNe are treated as Ia SNe or the magnitudes
are not modified does not have a significant effect. 
The corrections to the magnitudes are negligible compared to the
magnitude difference between the detected Ia and Ibc SNe.

\subsection{Colors}

From the sample of SNe compiled in Table~\ref{tab2}, we find that
the $B_{max}-V_{max}$ colors of Ibc SNe have a tendency to have
redder colors than Ia SNe, which typically have $B_{max}-V_{max}\sim 1$. 
However, the slope of the relation
of $B-V$ color with time from maximum is similar, as illustrated in 
Figure~15 of \citet{Setal02} for SN1999ee (Ia) and SN1999ex (Ib/c).
If the maximum was not observed, a shift of $\sim 15$ days would
place the two $B-V$ curves on top of one another.

If the $B_{max}-V_{max}$ color is correctly measured through
an accurate determination of the date at maximum, then an 
interloping Ibc SN will appear redder than the average Ia SN. If this
Ibc SN has a large decline parameter, it will be ``expected'' to be
fainter, and fainter Ia SNe are redder at maximum 
(see Figure~1 of Riess, Press, \& Kirshner (1996) for examples).
However, if this Ibc SN has a small decline parameter, it will be
``expected'' to be brighter, and brighter SNe are bluer. Therefore,
the red color will be interpreted as due to extinction. 

As an example, let us consider SN 1998bw \citep{Galamaetal98}, 
the first confirmed supernova/GRB. This supernova had a decline
parameter $\Delta m_{15}=1.0$,   
$B_{max}-V_{max}$ of 0.4 magnitudes, and M$_{B_{max}}=-18.88$. If the 
intrinsic color was
assumed to be $B_{max}-V_{max}=0$, then $E(B-V)=0.4$, $A_{V}=1.24$, and
$A_{B}=1.64$. The observed apparent rest-frame $B$ magnitude would be corrected by 
approximately 1.6 magnitudes, and the derived distance modulus would be 
too large by more than 1 magnitude. Put another way, this object would
appear 1 magnitude brighter than an accurately corrected Ia SN.

This effect should act in a random way, unless the peak magnitudes,
colors, and light curve shapes of Ibc SNe are correlated. With the
current available sample in the literature, we are unable to determine 
the relationship between
these parameters.
However, we note that some type Ia SNe have $B_{max}-V_{max}$ 
colors similar to those
of Ibc SNe, for example, 1990Y, 1992K, and 1993H \citep{Hetal96}.
If these or similar lightcurves are used as templates in the fitting 
procedure, then the argument in the previous paragraphs is negated.
For our example Ic SN 1998bw, the ``corrected'' mag would be approximately
0.6 magnitudes too faint.
In this case, the effect of intrinsic $B_{max}-V_{max}$ colors, which 
do appear to 
be different for Ia and Ibc SNe, can only be correctly assessed by 
testing the template fitting codes of individual surveys. We
urge investigators to consider testing the robustness of their
photometric and spectroscopic fitting to Ibc interlopers.

\subsection{SFR(z), Ibc/Ia(z)}

\begin{figure}
\includegraphics[scale=0.5]{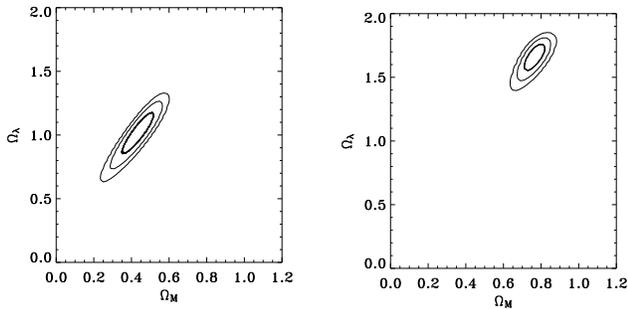}
\caption{Including a changing star formation rate with redshift with a single
gaussian Ibc magnitude distribution. In (a),
the rate of Type~Ibc to Ia SNe follows the dependence $0.05(1+z)$, while 
in (b) it varies as $0.1(1+z)$, the same redshift dependence but with a 
different normalization. The number of detected Ia and Ibc SNe was
(a) Ia$=318$, Ibc$=20$, and (b) Ia$=287$, Ibc$=41$. The minimum
$\chi^{2}$/N is (a) 1.57 and (b) 2.27.
  \label{sfrz}}
\end{figure}

\begin{figure}
\includegraphics[scale=0.5]{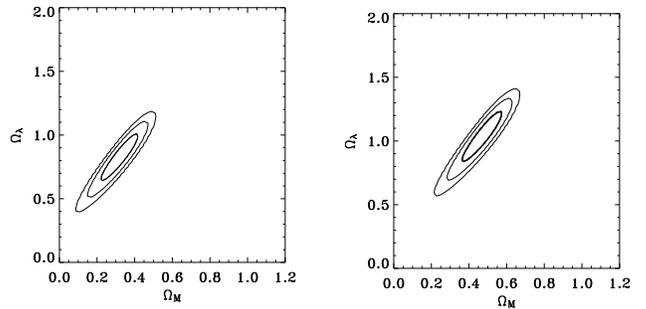}
\caption{Including a changing star formation rate with redshift with a 
two gaussian Ibc magnitude distribution. The Ia distribution is 
identical to that in Figure~\ref{sfrz}. In (a),
the rate of Type~Ibc to Ia SNe follows the dependence $0.05(1+z)$, while 
in (b) it varies as $0.1(1+z)$, the same redshift dependence but with a 
different normalization. The number of detected Ia and Ibc SNe was
(a) Ia$=318$, Ibc$=15$, and (b) Ia$=287$, Ibc$=32$. The minimum
$\chi^{2}$/N is (a) 1.21 and (b) 1.52.
  \label{sfrz_2}}
\end{figure}

We also investigated a Ibc/Ia ratio that changes with redshift. 
In Figure~\ref{sfrz} we present examples of including
a dependence of the Ibc/Ia ratio with redshift, mimicking a 
dependence on the star formation rate. Due to the severe
effect of Ibc contamination on the probability curves even 
for small levels of contamination, we were forced to keep the
relationship shallow ($1+z$) and with a small Ibc/Ia normalization 
ratio ($\le 0.1$). In Figure~\ref{sfrz}, the minimum reduced $\chi^{2}$
was 1.5. However, for the case shown in 
Figure~\ref{sfrz_2}, the minimum reduced $\chi^{2}$ was 2.6, and
the simple calculation of probability curves we have shown is not
valid. A more sophisticated treatment using the estimated Ia and non-Ia 
SNe rates with redshift from Dahl\'{e}n et~al. (2004) could be performed,
but the measurements are so sensitive to Ibc contamination that a 
more complex treatment is unwarranted.

%\clearpage

%\clearpage
%\clearpage

\subsection{Comparison to the $SNAP$ Redshift Distribution}

\citet{Kimetal04} reviewed systematic uncertainties 
in SNe studies to determine the cosmological parameters. In their
study, a fiducial number-redshift distribution for SNe was used, 
an observational scatter of 0.1 magnitudes, and a dispersion about
the corrected Ia peak magnitudes of 0.1 magnitudes.
We would like to use the same SNe-redshift distribution to see what
effect Ibc contamination would have. However, including the
redshift distribution of detected SNe is problematic, as instead of
specifying detected SNe, we assign an input distribution, then 
``observe'' those above our apparent magnitude limit. To accurately
determine the effects of Ibc contamination, an apparent magnitude limit
must be included, thus some ``observation'' must be simulated.
To match the $SNAP$ fiducial distribution, we tailor the 
input distribution such 
that the ``observed'' distribution is sufficiently similar to
the fiducial $SNAP$ distribution. 

\begin{figure*}
\begin{center}
  \includegraphics[width=15cm]{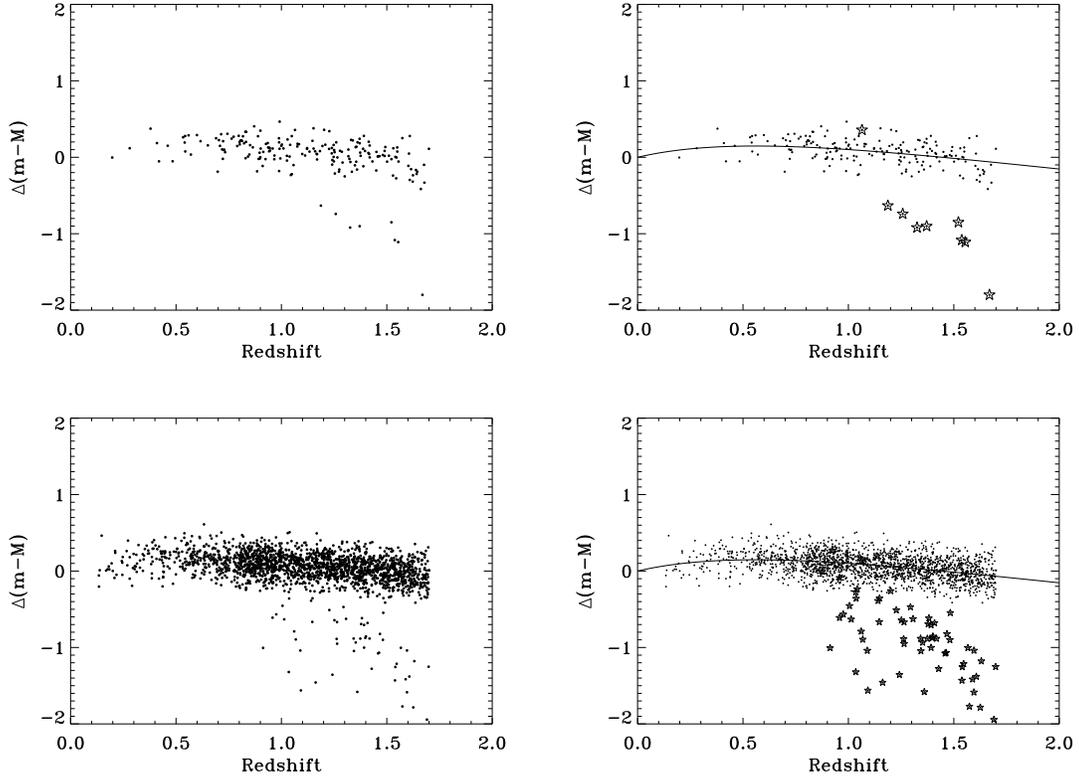}
  \caption{Observational scatter of 0.1 magnitudes. Top panels: 188 detected Ia, 9 Ibc, bottom panels: 1922 detected Ia, 64 Ibc. Stars in the right panels indicate Ibc SNe.
    \label{snap_smsc}}
\end{center}
\end{figure*}

We attempt to reproduce the $SNAP$ fiducial distribution's
general increase in numbers from 
$z=0.1-0.8$, the peak near z=0.8, and the 
slope of the decline in numbers from $z=0.8-1.7$, $dlogN(z)/dz = -0.4$.
The solution of input to detected numbers is not general, but depends
on the detection limit. For $m_{B}=25.$, we find a reasonable agreement
with an input of two power law distributions, 700 SNe between $z=0.1-1.0$
as $N \propto z^{2}$, and
10000 between $z=0.8-1.7$ as $N \propto z^{5}$. We then input 1000 Ibc SNe 
between $z=0.8-1.7$ as $N \propto z^{5}$, the same 
relationship as for the high redshift Ia population. Thus, we include an 
underlying ratio of 0.1 for Ibc/Ia SNe. An example of a
final detected redshift distribution is shown in Figure~\ref{snap_smsc}(a),
and the $\Delta (m-M)$ vs. z distribution in Figure~\ref{snap_smsc}(b). We 
emphasize that this is one run of a Monte Carlo code, and as in every random
process, each individual answer is different. 

\begin{figure*}
\begin{center}
  \includegraphics[width=15cm]{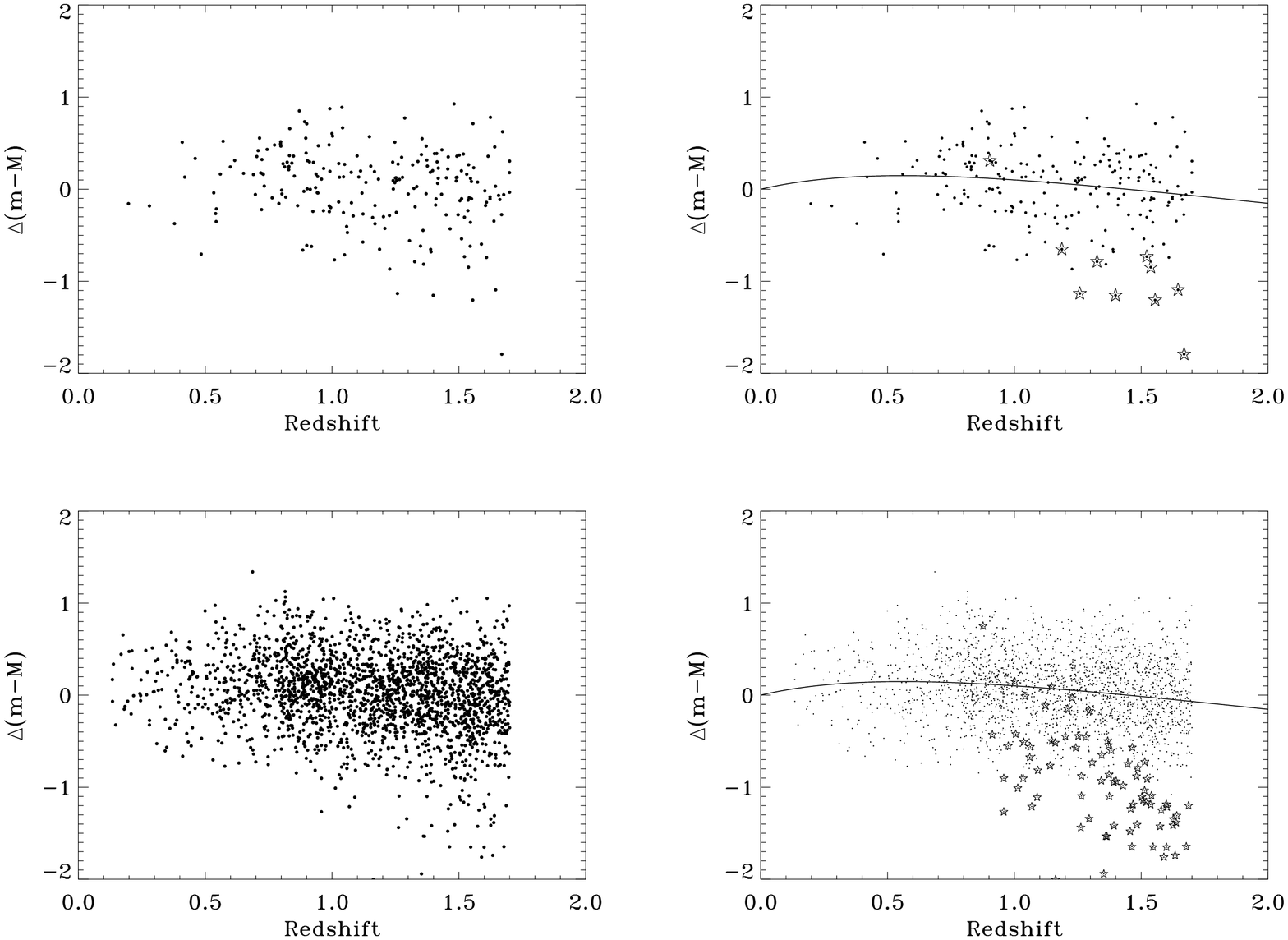}
  \caption{Observational scatter of 0.35 magnitudes.
top panels: 207 detected Ia, 10 Ibc, bottom panels: 2029 detected Ia, 80 Ibc. 
Stars in the right panels indicate Ibc SNe.
    \label{snap_lgsc}}
\end{center}
\end{figure*}

An important issue
is clearly illustrated in Figure~\ref{snap_smsc}; if the
observational scatter and the dispersion 
about the corrected peak magnitudes is low, 0.1 magnitudes for 
each in this simulation, then Ibc SNe are easily distinguished. 
Even when the number of SNe is low, the bright Ibc SNe that bias 
measurements of the cosmological parameters are easily isolated.
On the other hand, when the observational scatter is large, large
numbers of SNe are needed to identify outliers, in this case Ibc SNe. 
This is illustrated in Figure~\ref{snap_lgsc}.

{\it If} such a low scatter and corrected dispersion is feasible, 
then the need for spectra of SNe candidates to determine Ia-ness is
eliminated. Depending on the distribution of SNe, such a survey 
could possibly rely instead on rejecting $3\sigma$ outliers
from the mean or median distance modulus in a given redshift bin.
Even if the observational scatter is $\sim 0.3$, if the numbers are large, then 
$\sigma$-clipping is still a valid option to remove outliers, however,
the constraints on the cosmological parameters will not improve, 
as the probability curves will not shrink. 
The important question becomes, is such precision feasible? The answer to
this question is best left to survey designers.

%\clearpage
%\clearpage

%\clearpage

\section{On the Likelihood of I\lowercase{bc} Contamination}

As the progenitors of Type Ibc SNe are massive stars, we can 
expect that the majority have higher
extinctions than Ia SNe and occur in gas-rich galaxies. Thus,
excluding SNe with high derived $A_{V}$ values and those occuring in
late-type host galaxies should reject more Ibc SNe than Ia and
reduce contamination. However, the SN rate in different galaxy types 
as a function of redshift is not known. Although the 
local Ibc/Ia rate is $\sim 0.1$, we can expect this to increase
to $0.5-1.0$ around $z=1$. Assuming no evolution in the luminosity
functions, this means $1-3$ {\it bright} Ibc SNe for every 10 Ia.
For Ibc contamination to be a significant concern, 
$20-50$\% of these {\it bright} Ibc SNe must fail to be rejected 
through the possible combinations of light curve fitting,
spectral-typing, host galaxy morphology, and extinction.
The likelihood of this could be determined through detailed simulations
using Ibc templates (e.g. 1999bw, 1992ar) and routines such as MLCS2k2 
\citep{Retal04b}, or BATM \citep{Tonetal03}.

\section{Conclusions}

We find that even small contamination fractions (5\%) may bias the 
measurement of $\Omega_{M}$ and $\Omega_{\Lambda}$. However, we 
have also illustrated 
the ease with which Ibc SNe can be identified when the observational
error and the dispersion about the corrected Ia peak magnitudes is low.
Based on the calculations presented here, the straightforward
effect of contamination 
from Type~Ibc SN in intermediate-to-high redshift SN samples is 
to reduce the $\chi^{2}$
goodness-of-fit to the true cosmological parameters. This manifests itself
as an increase in the minimum value of the reduced $\chi^{2}$ grid.
For a flat universe with $\Omega_{M}=0.27, \Omega_{\Lambda}=0.73$
even small numbers (5\%) of Type~Ibc SNe may push both values upwards. 
Erroneously small 
``probability'' curves can be generated by assuming the model is a 
good fit and that the minimum reduced $\chi^{2}$ is close enough to 1
for a meaningful conversion to a probability grid. However, 
if the observational errors are large,
it is difficult to distinguish between
a minimum reduced $\chi^{2}$ that is larger due to a mismatch of the model
and the observations due to contamination or from simple stochastic effects.

If the observational errors are large enough ($> 0.3$), then 
a small contamination fraction (5\%)
by Ibc SNe can shift the location of the $\chi^{2}$ minimum without
making $\chi^{2}$ itself large. In such a case the error contours will be 
large,
so while the contours are not strictly statistically correct, they
are still meaningful in the sense that the underlying cosmological
parameters have close to a 99\% chance of being within the $2\sigma$
contours. 
If the observational error approaches the scatter in the
corrected 'standard candle' Ia magnitudes, the errors bars will shrink.
However, with the same amount of Ibc contamination, as the observational 
error steadily decreases, the reduced $\chi^{2}$ 
becomes steadily greater than 1. Then the conversion from a $\chi^{2}$ grid
to probabilities is not valid. One may assume that the random errors are
underestimated and renormalize; the error contours may then become 
large enough to encompass the
underlying $\Omega_{M},\Omega_{\Lambda}$ values. But if this precision
is achieved, outliers should already be readily discriminated in the
$\Delta (m-M)$ vs. $z$ plane, and a simple $3-5 \sigma$ clipping routine 
may suffice. Simply having large numbers of SNe with the precision of 
current surveys would also allow one to use $\sigma$-clipping to remove
Ibc outliers, however, this will not achieve 
significantly greater precision in the measurement of cosmological 
parameters.

%% In this section, we use  the \subsection command to set off
%% a subsection.  \footnote is used to insert a footnote to the text.


%% This section also includes several of the displayed math environments
%% mentioned in the Author Guide.

\acknowledgments
The author is pleased to acknowledge the anonymous referee for comments 
and suggestions which improved the content of the paper.  Thank you to 
T. Matheson for providing lightcurve data
and reference pointers for Ib/c SNe, and A. Riess for useful comments on
how SN typing is performed. The author would also like to thank B. Leibundgut 
for comments on an
early version of this manuscript, J. Tonry for providing
his code for perusal, and especially hearty thank-yous to 
T.~Puzia, J.~Blakeslee, and M.~Livio for critical readings and comments. 
This work was supported in part by generous scientific
freedom granted by H. Ford, for which the author is grateful.


\begin{thebibliography}{}

\bibitem[Barris et al.(2004)]{Barrisetal04} Barris, B.~J., et al.\ 
2004, \apj, 602, 571 

\bibitem[Blakeslee et al.(2003)]{Blaketal03} Blakeslee, J.~P.~et 
al.\ 2003, \apj, 589, 693 

\bibitem[Bouwens et al.(2003)]{Bouwetal03} Bouwens, R.~J.~et al.\ 
2003, \apj, 595, 589 

\bibitem[Clocchiatti et al.(1996)]{Cloetal96} Clocchiatti, A., et 
al.\ 1996, \aj, 111, 1286 

\bibitem[Clocchiatti et al.(1997)]{Cloetal97} Clocchiatti, A., et 
al.\ 1997, \apj, 483, 675 

\bibitem[Clocchiatti et al.(2000)]{Cloetal00} Clocchiatti, A., et 
al.\ 2000, \apj, 529, 661 

\bibitem[Clocchiatti et al.(2001)]{Cloetal01} Clocchiatti, A., et 
al.\ 2001, \apj, 553, 886 

\bibitem[Coil et al.(2000)]{Cetal00} Coil, A.~L., et al.\ 2000, 
\apjl, 544, L111 

\bibitem[Dahl\'{e}n et~al.(2004)]{Dahlenetal04} Dahl\'{e}n, T., et al. \
2004, \apj, in press

\bibitem[Galama et al.(1998)]{Galamaetal98} Galama, T.~J., et al.\ 
1998, \nat, 395, 670 

\bibitem[Gal-Yam et al. (2004)]{GYetal04} Gal-Yam, A., Poznanski, D., Maoz, 
D., Filippenko, A.~V., \& Foley, R.~J 2004, PASP, in press

\bibitem[Hamuy et al.(1996)]{Hetal96} Hamuy, M., Phillips, 
M.~M., Suntzeff, N.~B., Schommer, R.~A., Maza, J., \& Aviles, R.\ 1996, 
\aj, 112, 2391 

\bibitem[Kim et al.(2004)]{Kimetal04} Kim, 
A.~G., Linder, E.~V., Miquel, R., \& Mostek, N.\ 2004, \mnras, 347, 909 

\bibitem[Leibundgut(2001)]{Leib01} Leibundgut, B.\ 2001, 
\araa, 39, 67 

\bibitem[Lilly et al.(1996)]{Letal96} 
Lilly, S.~J., Le Fevre, O., Hammer, F., \& Crampton, D.\ 1996, \apjl, 460, 
L1 

\bibitem[Madau, Pozzetti, \& Dickinson(1998)]{Madetal98} Madau, 
P., Pozzetti, L., \& Dickinson, M.\ 1998, \apj, 498, 106 

\bibitem[Matheson et al.(2001)]{Metal01} Matheson, T., 
Filippenko, A.~V., Li, W., Leonard, D.~C., \& Shields, J.~C.\ 2001, \aj, 
121, 1648 

\bibitem[Nugent, Kim, \& Perlmutter(2002)]{NKP02} Nugent, P., 
Kim, A., \& Perlmutter, S.\ 2002, \pasp, 114, 803 

\bibitem[Perlmutter et al.(1999)]{Perletal99} Perlmutter, S.~et 
al.\ 1999, \apj, 517, 565 

\bibitem[Phillips(1993)]{P93} Phillips, M.~M.\ 1993, \apjl, 
413, L105 

\bibitem[Richardson et al.(2002)]{Richetal02} Richardson, D., 
Branch, D., Casebeer, D., Millard, J., Thomas, R.~C., \& Baron, E.\ 2002, 
\aj, 123, 745 

\bibitem[Richmond et al.(1994)]{Retal94} Richmond, M.~W., 
Treffers, R.~R., Filippenko, A.~V., Paik, Y., Leibundgut, B., Schulman, E., 
\& Cox, C.~V.\ 1994, \aj, 107, 1022

\bibitem[Richmond et al.(1996)]{Retal96} Richmond, M.~W., et 
al.\ 1996, \aj, 111, 327 

\bibitem[Richmond, Filippenko, \& Galisky(1998)]{RFG98} 
Richmond, M.~W., Filippenko, A.~V., \& Galisky, J.\ 1998, \pasp, 110, 553 

\bibitem[Riess, Press, \& Kirshner(1996)]{RPK96} Riess, 
A.~G., Press, W.~H., \& Kirshner, R.~P.\ 1996, \apj, 473, 88 

\bibitem[Riess et al.(1998)]{Riessetal98} Riess, A.~G.~et al.\ 
1998, \aj, 116, 1009 

\bibitem[Riess et al.(2004a)]{Retal04a} Riess, A.~G.~et al.\ 
2004a, \apjl, 600, L163 

\bibitem[Riess et al.(2004b)]{Retal04b} Riess, A.~G., et al.\ 
2004b, \apj, 607, 665 


\bibitem[Steidel et al.(1999)]{Steidetal99} Steidel, C.~C., 
Adelberger, K.~L., Giavalisco, M., Dickinson, M., \& Pettini, M.\ 1999, 
\apj, 519, 1 

\bibitem[Stritzinger et al.(2002)]{Setal02} Stritzinger, M., et 
al.\ 2002, \aj, 124, 2100 

\bibitem[Strolger et~al.(2004)]{Strolgeretal04} Strolger, L~.G., et al. \
2004, \apj, in press

\bibitem[Thompson, Weymann, \& Storrie-Lombardi(2001)]{Thometal01} 
Thompson, R.~I., Weymann, R.~J., \& Storrie-Lombardi, L.~J.\ 2001, \apj, 
546, 694 

\bibitem[Tonry et al.(2003)]{Tonetal03} Tonry, J.~L.~et al.\ 
2003, \apj, 594, 1 

\end{thebibliography}
\end{document}